\documentstyle[prl,aps,twocolumn,psfig]{revtex}
\begin{document}
\preprint{}
\draft
%
%
\title{Chaotic properties of quantum many-body systems 
in the thermodynamic limit}

\author{Giovanni Jona-Lasinio\cite{AJ} and Carlo Presilla\cite{AP}}
\address{Dipartimento di Fisica, Universit\`a di Roma ``La Sapienza,''
and INFN, Sezione di Roma,\\
Piazzale A. Moro 2, 00185 Roma, Italy}
\date{cond-mat/9601056 submitted to Phys. Rev. Lett.}
\maketitle
%
%
\begin{abstract}
By using numerical simulations, we investigate the dynamics
of a quantum system of interacting bosons.
We find an increase of properly defined mixing properties when the
number of particles increases at constant density or the interaction
strength drives the system away from integrability.
A correspondence with the dynamical chaoticity of an associated $c$-number 
system is then used to infer properties of the quantum system 
in the thermodynamic limit.
\end{abstract}
%
%
%
%
Classical Hamiltonian systems are usually termed chaotic 
if their trajectories show local exponential instability, i.e., 
a positive Lyapunov exponent \cite{GUTZWILLER}.
This definition reflects the generally nonlinear character of the 
differential equations of the classical motion.
Here we will refer to this situation as {\em dynamical chaos}.

At quantum level, every system is described by a linear Schr\"odinger 
equation and dynamical chaos is not possible.
One can resort to a definition of quantum chaos based on the correspondence
principle.
It is often assumed, but not rigorously proved, that classically chaotic
systems give rise to quantum mechanical spectra whose statistical
properties are well described by random matrix theory \cite{BGS}.
Indeed, a large class of numerical examples \cite{BOHIGAS}
and recent theoretical work \cite{AASA}, indicate that the nearest neighbor
level spacing (NNLS) distribution of systems which are classically chaotic is
well approximated by a Wigner distribution \cite{GUTZWILLER}. 

The above statements refer to confined systems with a finite number of 
degrees of freedom. 
Quantum mechanically these systems are characterized by a 
discrete spectrum.
The situation is different if we consider a many-body system in the 
thermodynamic limit, i.e., when the number $N$ of particles tends
to infinity at constant density.
In this limit the spectrum is, in general, continuous and true chaotic
phenomena are not excluded \cite{CHIRIKOV,JPC}.

One should state clearly from the outset that in the thermodynamic
limit chaotic behavior, in the sense that the system is mixing, can appear 
through a mechanism which has nothing to do with the nonlinearity of the
interaction but is connected with the possibility of transforming 
space chaos into time chaos.
This is true both at classical and quantum level as it is illustrated in 
the case of an infinite system of linearly interacting oscillators 
classical \cite{LL}, or quantum \cite{GRAMAR}, and in the case of
a gas of noninteracting particles classical \cite{SINAI}, 
or quantum \cite{LENCI}.
To clarify the point in question let us consider the case of a 
one-dimensional lattice of classical harmonic oscillators coupled in such 
a way that it may be considered as the discretization of the wave 
equation in one space-dimension.
As it is well known the solutions of this equation depend on the
combination $x \pm t$, 
where $x$ and $t$ are the space and time coordinates, respectively.
As a consequence, if the initial condition is a realization of a 
mixing stochastic process in space this is transformed by the dynamics 
into a mixing stochastic process in time at any point in space. 
If the system is in equilibrium at some temperature $T$,
the initial conditions to be considered are typical realizations of the
stochastic process corresponding to the equilibrium Gibbs measure.
We shall call the chaotic behavior of noninteracting or linearly
interacting many-body systems {\em kinematical chaos}.
Clearly, in the cases considered in \cite{GRAMAR,LENCI} Lyapunov
exponents are zero because the dynamics in the thermodynamic
limit is the limit of the finite-dimensional dynamics.

In this paper we want to investigate what happens when a nonlinear
interaction is switched on, i.e., the Hamiltonian describing the system
is not quadratic.
We begin by recalling the general definition of mixing given in 
\cite{GRAMAR,LENCI}.
Let us consider $N$ interacting particles in a volume $V \subset \bbox{R}^d$
described by the Hamiltonian $\hat{H}$ and let $\hat{A}$ and $\hat{B}$ 
be two local observables.
We shall say that the system has the property of quantum mixing in the
thermodynamic limit if the following holds:
\begin{equation}
\lim_{t \to \infty} \lim_{{N,V \to \infty} \atop {N/V \to \rho}}
\langle \hat{A}(t)~\hat{B} \rangle = 
\lim_{{N,V \to \infty} \atop {N/V \to \rho}} \langle \hat{A} \rangle ~
\lim_{{N,V \to \infty} \atop {N/V \to \rho}} \langle \hat{B} \rangle 
\label{MIXDEF}
\end{equation}
where 
\begin{eqnarray}
\langle ~\ldots~ \rangle =
{\mbox{Tr}~\ldots ~e^{- \hat{H}/k_BT} \over \mbox{Tr}~
e^{- \hat{H}/k_BT} },
~~~~~~\hat{A}(t)=
e^{{i\over \hbar} \hat{H}t} \hat{A} e^{-{i\over \hbar} \hat{H}t}.
\nonumber
\end{eqnarray}
As a rule, the limits on the l.h.s. of (\ref{MIXDEF}) must be taken in the
order indicated.
However, for the systems discussed in \cite{GRAMAR,LENCI} the limits
can actually be inverted and we shall assume that this applies also 
to the nonlinearly interacting systems considered in this paper.

Since, even if the interaction is nonlinear, the finite-dimensional dynamics
is quasi-periodic, Lyapunov exponents are zero for any finite $N$ and,
therefore, also in the thermodynamic limit \cite{GOLDSTEIN}.
We expect, however, an influence of the nonlinearity on the mixing
properties of the system.
In particular, we expect that the strength of the nonlinearity
will affect the rate of convergence of the $t$ limit in (\ref{MIXDEF})
once the thermodynamic limit has been taken.
If $N$ is finite, as it will be the case in computer simulations, 
we expect the difference
\begin{equation}
\langle \hat{A}(t) \hat{B} \rangle - \langle \hat{A} \rangle 
\langle \hat{B} \rangle
\label{ABAB}
\end{equation}
to oscillate in time with an averaged amplitude (see Eq. (\ref{MIXIND})) 
which decreases when the interaction drives the system away from integrability.
This amplitude should decrease also when the interaction strength is kept 
fixed and $N$ increases with $N/V$ constant.

We consider a system of $N$ 
spinless bosons of charge $q$ moving in a one-dimensional lattice 
with $L$ sites and described by the Hamiltonian
\begin{eqnarray}
\hat{H} &=& \sum_{j=1}^L \left [ \alpha_j \ \hat{a}_j^{\dagger} \hat{a}_j - 
\beta_j \left( e^{i \theta} \ \hat{a}_{j+1}^{\dagger}\hat{a}_{j}
            + e^{-i \theta} \ \hat{a}_j^{\dagger} \hat{a}_{j+1} \right) \right] 
            \nonumber \\
&& + \sum_{j=1}^L  \ \gamma_j \ \hat{a}_{j}^{\dagger} 
\hat{a}_{j}^{\dagger}  \hat{a}_{j} \hat{a}_{j} ,
\label{H}
\end{eqnarray}
where index correspondence $j \pm L=j$ is assumed.
The operator $\hat{a}_j^{\dagger}$ creates a boson in 
the site $j$ and $\alpha_j$, $\beta_j$, and $\gamma_j$ are 
the site, hopping, and interaction energies, respectively.
Dirichlet or periodic boundary conditions can be chosen. 
In the first case the sites lie on a segment and we put 
$\beta_L=0$ and $\theta=0$.
In the second case the system represents a ring 
threaded by a line of magnetic flux $\phi$. 
The phase factors are $\theta=2 \pi \phi / \phi_0 L$, 
where $\phi_0=hc/q$ is the flux quantum.
The system (\ref{H}) has wide interest.
Its time-dependent mean-field approximations have applications to  
molecular dynamics and nonlinear optics \cite{EILBECK}
and to electron transport in heterostructures \cite{PJC}.

For finite $L$ and $N$ the dimension of the Fock space spanned by the 
system (\ref{H}) is finite and given by 
\begin{equation}
D = \frac{(N+L-1)!}{N! \ (L-1)!} .
\end{equation}
The $D$-dimensional matrix representing 
the Hamiltonian (\ref{H}) in the base of the Fock states 
$| n_1^i \cdots n_L^i \rangle$, $i=1,\cdots,D$, 
where $n_j^i$ is the occupation number of the $j$th site in the $i$th Fock
state and $\sum_{j=1}^L n_j^i=N$, can be diagonalized by standard numerical
methods with negligible errors.

We have calculated the quantity (\ref{ABAB}) for different local operators 
$\hat{A}$ and $\hat{B}$ and for $N \leq 7$ with $N/L=1$.
Figure 1 shows typical results obtained at zero temperature for 
$\hat{A}=\hat{B}=\hat{a}_{k+1}^{\dagger} \hat{a}_k +
\hat{a}_{k}^{\dagger} \hat{a}_{k+1}$ with $k=3$.
For simplicity, in the numerical simulation we put $\alpha$, 
$\beta$ and $\gamma$ independent of the site $j$.
The number of particles and sites considered, $N=L=5$ and $N=L=7$, 
may look very small but one has to remember that the complexity of the 
system is given by the Fock dimension $D$ which is 126 and 1716, respectively.
Up to such values of $D$ we observe that the amplitude of the oscillations 
of (\ref{ABAB}) decreases in presence of nonlinear interaction.
By comparing Figs. 1 a and b, we have also evidence of a decrease of these 
oscillations with increasing $N$ at constant density.

To make the above discussion quantitative we measure the oscillating behavior
of (\ref{ABAB}) by introducing the following indicator
\begin{equation}
\kappa_{\hat{A} \hat{B}} = \lim_{t \to \infty} 
\sqrt{ {1 \over t} \int_0^t dt'~
\left| \langle \hat{A}(t')~\hat{B} \rangle - 
\langle \hat{A} \rangle ~ \langle \hat{B} \rangle \right|^2 }.
\label{MIXIND}
\end{equation}
Mixing implies that $\kappa_{\hat{A} \hat{B}} = 0$.
If $\hat{A}$ or $\hat{B}$ commutes with the Hamiltonian $\hat{H}$,
$\langle \hat{A}(t) \hat{B} \rangle - \langle \hat{A} \rangle 
\langle \hat{B} \rangle$ is identically zero.
The limit (\ref{MIXIND}) can be evaluated exactly and in Fig. 2 we show 
its value for the same observables of Fig. 1 as a function of the 
ratio $\gamma/\beta$. 
For fixed $\alpha$, $\kappa_{\hat{A} \hat{B}}$ depends only on this ratio.
Since the system considered is integrable for $\gamma=0$ and $\beta=0$
we expect a minimum of $\kappa_{\hat{A} \hat{B}}$,
that is a maximal chaoticity, between these two limits.
For $\alpha=0$ and $N/L=1$, this minimum should take place at 
$\gamma/\beta \sim 1$.
The results shown in Fig. 2 confirm this expectation as well the decrease
of  $\kappa_{\hat{A} \hat{B}}$ when $N$ increases at constant density.
Similar results are obtained at finite temperature.

There is a natural $c$-number system associated to a system of bosons
like the one we consider.
This is obtained by constructing a mean-field approximation which, when
quantized, reproduces the exact quantum equation in the second quantization
formalism \cite{CJP}.
We now provide evidence that there exists a strict correspondence between
the dynamics of this $c$-number system and that of the quantum system
by evaluating the maximal Lyapunov 
exponent of the mean-field dynamics as a function of the interaction 
strength $\gamma$.
Nonlinear mean-field equations for the system (\ref{H}) can be written 
as \cite{BR}
\begin{eqnarray}
i \hbar {d \over dt} z_j(t) &=& 
\left[ \alpha_j + 2 (N-1) \gamma_j |z_j(t)|^2 \right] z_j(t) 
\nonumber \\
&& - \beta_{j-1} e^{i\theta}  z_{j-1}(t) 
   - \beta_{j} e^{-i\theta} z_{j+1}(t), 
\label{EZ}
\end{eqnarray}
where $z_j(t)$ is the amplitude of the mean field in the site $j$.
Conservation of the single-particle probability
\begin{equation}
\sum_{j=1}^L |z_j(t)|^2 = 1 
\label{CONSN}
\end{equation}
and of the single-particle energy
\begin{eqnarray}
{\cal E}[z,z^*] &=& \sum_{j=1}^L \Bigl\{
\alpha_j |z_j(t)|^2 + (N-1) \gamma_j |z_j(t)|^4  
\nonumber \\
&& - \left[ \beta_{j-1} e^{i\theta} z_{j-1}(t)
   + \beta_j e^{-i\theta} z_{j+1}(t) \right] z_j^*(t) \Bigr\} 
\label{CONSE}
\end{eqnarray}
are crucial constraints for a correct numerical simulation of (\ref{EZ}).
The corresponding maximal Lyapunov exponent $\lambda$ can be then 
numerically evaluated with negligible errors \cite{CJP}.
In order to use a dimensionless quantity we consider the rescaled exponent
$\lambda \hbar/\gamma$ when the ratio $\gamma/\beta$ is varied
at fixed $\beta$.
In fact, $\lambda$ depends on both $\beta$ and $\gamma/\beta$. 
The curve in Fig. 3 shows that $\lambda \hbar/\gamma$ has a pronounced 
maximum in the same region where $\kappa_{\hat{A} \hat{B}}$ has a minimum.
This means that the tendency to chaoticity of the finite quantum system and 
the chaoticity of its $c$-number 
counterpart have the same qualitative behavior away from 
integrability points.
This aspect can be analyzed in greater detail and will be discussed in
a subsequent publication.

The numerical study of the quantum system (\ref{MIXDEF}) becomes prohibitive 
for values of $D$ larger than a few thousands. 
To get an idea of what happens when we increase further the number of
particles, we make the reasonable hypothesis that the chaotic behavior 
increases if the same happens for the corresponding $c$-number system.
The numerical evaluation of the maximal Lyapunov exponent of the mean-field
dynamics is feasible also for large values of $N$ and $L$.
We now show that the chaotic behavior of the mean-field evolution 
of our system increases monotonically and eventually becomes constant 
when the thermodynamic limit is approached.
Figure 4 displays the behavior of the maximal Lyapunov exponent
of (\ref{EZ}) up to $N=2000$ for $N=L$ and $\gamma \simeq \beta$, 
that is in the region of maximal chaoticity. 
We have a satisfactory evidence that the maximal Lyapunov exponent 
reaches a limiting value which we assume to characterize the dynamics
in the thermodynamic limit.

On the basis of these results we conclude that the maximal chaoticity 
of the quantum system should also increase steadily until the limit
$\kappa_{\hat{A}\hat{B}}=0$ is reached.

The approach that we have developed in this paper is clearly applicable to
the study of the chaotic properties of any quantum system and is
complementary to the usual search of quantum signatures of classical chaos.

%
%

We thank J. Bellissard, S. Graffi, and A. Martinez for very
enlightening discussions.
We are very grateful to Y. Avron, V. Bach, and R. Seiler
for the warm hospitality at the Erwin Schr\"odinger Institute 
in Wien where part of this work was done. 
Partial support of INFN, Iniziativa Specifica RM11, is acknowledged.
%
%

%
\newpage
\begin{figure}   
\centerline{\hbox{\psfig{figure=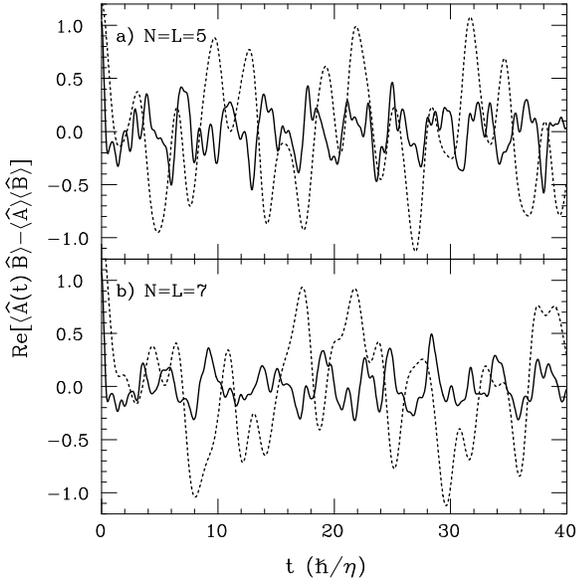,width=11.0cm,angle=90}}}
\caption{Real part of (\protect\ref{ABAB}) as a function of time
for $N=L=5$ (a) and $N=L=7$ (b) at zero temperature.
The system has periodic boundary conditions with $\phi/\phi_0=0.3$, 
$\alpha=0$, $\beta=\eta$ and $\gamma=\eta$ (solid line)
or $\gamma=0$ (dashed line).
The local operators considered are 
$\hat{A}=\hat{B}=\hat{a}_{k+1}^{\dagger} \hat{a}_k +
\hat{a}_{k}^{\dagger} \hat{a}_{k+1}$ with $k=3$.}
\label{FIG1}
\end{figure}

\begin{figure}   
\centerline{\hbox{\psfig{figure=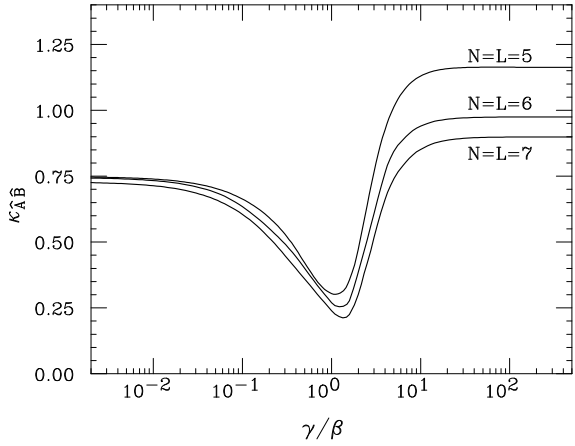,width=9.0cm,angle=90}}}
\caption{
Mixing indicator (\protect\ref{MIXIND}) at zero temperature for the system 
of Fig. 1 as a function of the ratio $\gamma/\beta$.}
\label{FIG2}
\end{figure}

\begin{figure}   
\centerline{\hbox{\psfig{figure=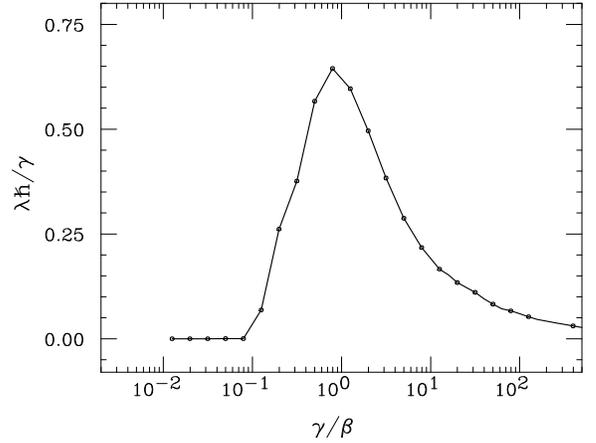,width=9.0cm,angle=90}}}
\caption{
Rescaled maximal Lyapunov exponent $\lambda \hbar /\gamma$ of the 
mean-field system (\protect\ref{EZ}) as a function of the ratio 
$\gamma/\beta$ for $\alpha=0$, $\beta=\eta$, $N=L=5$, $\phi/\phi_0=0.3$
and periodic boundary conditions.
The initial mean-field components $z_j(0)$ are arbitrary complex numbers 
with $|z_j(0)|^2=1/L$.}
\label{FIG3}
\end{figure}

\begin{figure}   
\centerline{\hbox{\psfig{figure=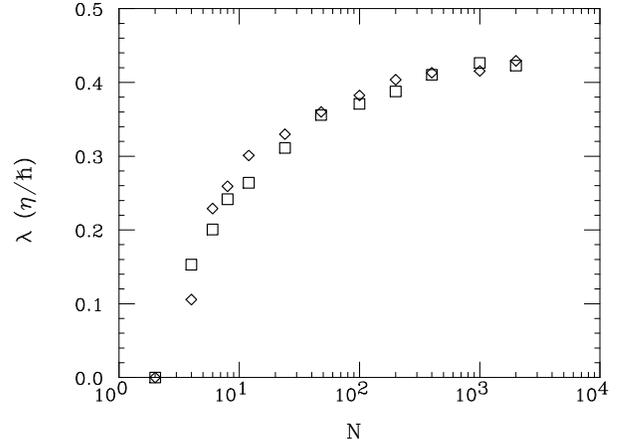,width=9.0cm,angle=90}}}
\caption{
Maximal Lyapunov exponent $\lambda$ of the mean-field 
system (\protect{\ref{EZ}}) as a function of the number $N$ of particles 
at constant density $N/L=1$. 
We have $\alpha=0$, $\beta=\eta$, and $\gamma = \eta L/(N-1)$ 
with $\phi/ \phi_0=0$ and Dirichlet boundary conditions (squares) 
and $\phi/ \phi_0=0.3$ and periodic boundary conditions (diamonds).
The initial mean-field components $z_j(0)$ are chosen in order to have 
the same single-particle energy for any value of $N$.}
\label{FIG4}
\end{figure}

\end{document}